\documentstyle[11pt,moriond,epsfig]{article}

\def\Journal#1#2#3#4{{#1} {\bf #2}, #3 (#4)}

\def\be{\begin{equation}}
\def\ee{\end{equation}}
\def\bea{\begin{eqnarray}}
\def\eea{\end{eqnarray}}

\begin{document}
\vspace*{4cm}
\title{THE VERY SMALL ARRAY: STATUS REPORT}

\author{ Michael E. JONES \& P. F. SCOTT }

\address{Cavendish Laboratory, Madingley Road, Cambridge CB3 0HE, UK.}

\maketitle\abstracts{We are contructing an interferometric telescope, the Very
Small Array, to study the cosmic microwave background on angular scales
0.2--$4.5^{\circ}$. The physical layout and electronic design of the telescope
are optimised to give maximum protection from systematic effects, while still
providing sufficient sensitivity to make high signal-to-noise images. A
prototype single baseline is currently being tested, with scientific results
expected during 2000.}

\section{Introduction}

It is widely accepted that the majority of the cosmological results that might
be obtained from the cosmic microwave background (CMB) will come from accurate
measurements of the CMB power spectrum over the region of the accoustic peaks,
that is in the range $100 < l < 2000$ (where $l$ is the spherical harmonic
multipole). We have therefore been constructing an instrument, the Very Small
Array (VSA), to measure the power spectrum (and provide images) in precisely
this angular range. We have argued previously\cite{jones} that interferometers
are very well suited for ground-based CMB measurements, given their relative
immunity to the atmosphere and other systematic effects compared with
switched-beam experiments. Here we will describe some of the more detailed
design considerations of the array, and review the progress of the project so
far.

\section{The VSA---vital statistics}

The VSA will consist of an array of 14 antennas, operating in the 26--36~GHz
band. The size of the antennas is determined by the largest angular scale we
wish to observe: $l = 100$ corresponds to an angular scale of $3^{\circ}.6$,
or an interferometer baseline of 16 wavelengths. Our antennas have a clear
aperture of 143~mm, ie about 15 wavelengths in the centre of the band. The
number of antennas is fixed simply by the cost and complexity of the whole
telescope---the more simultaneous baselines present, the better. In order to
measure the smallest angular scales ($l = 2000$ corresponds to a baseline of
$320 \lambda$) without compromising the fraction of the synthesised aperture
that is filled, we will also use alternative antennas about 2.25 time larger. 

Using the best available front-end amplifiers (a design kindly supplied by
NRAO), an observing bandwidth of 1.5~GHz (limited by bandwidth smearing), and
a site with good transparency and stability at 30~GHz (Teide Observatory,
Tenerife), we should obtain maps with $7\,\mu$K sensitivity per $30^{\prime}$
($12^{\prime}$) pixel over a $4^{\circ}.5$ ($2^{\circ}$) field in 300 hours
observation, using the smaller (larger) antennas.

\section{Design considerations}

\subsection{Tracking arrangement}

One of the main advantages of interferometers is {\em fringe rotation}. That
is, because the rotation of the earth constantly changes the relative path
from the source to each antenna, the complex visibility due to the source is
phase modulated at a rate that depends on the baseline vector and the
direction to the source. Synchronous detection at this frequency results in
all spurious signals, due to eg crosstalk, atmospheric emission, or bright
sources in far sidelobes, being attenuated by factors which can be very large
(easily 25~dB). However, this is only true if the antennas are mounted so that
the relative path does change, e.g. if the antennas are fixed to the ground
and track individually.

A conflicting requirement is that the antennas be very closely packed. The
sensitivity of a synthesised aperture is degraded compared to a single
aperture with the same beam by the fraction of the aperture that is
filled. For an array of $N$ antennas of size $d$, aperture efficiency $\eta$
and maximum baseline $D$, this factor is roughly $f = \eta N (D/d)^2$. It is
much easier to pack antennas very closely if they do not move relative to each
other.

We therefore need a design in which the antennas can be packed closely in two
dimensions, yet have a large change of relative path as they track. They must
also, of course, have excellent sidelobe performance (the Sun is about 90~dB
more powerful than the final noise level) and be free from crosstalk.

\begin{figure}
\epsfig{figure=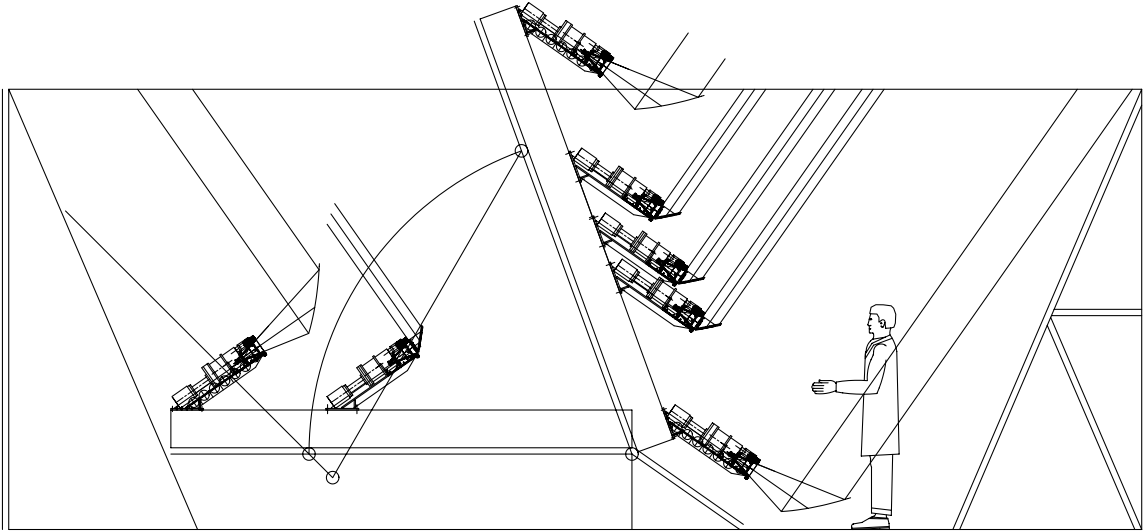,angle=90,width=6in}
\caption{Sketch of VSA layout showing both large and small antennas, and the
two extreme positions of the tip-table.\label{table}}
\end{figure}

The design we have selected (Fig. \ref{table}) achieves these goals by using
the the conical horn-reflector antenna (CHRA) design\cite{ghassan} already
used successfully on the Cosmic Anisotropy Telescope\cite{robson}, with a
modified mounting arrangement. The antennas are mounted on a table, with the
horn axis in a vertical N-S plane. Tracking in the E-W direction is obtained
by rotating the mirror about the optic axis of the horn; in the N-S direction,
by tipping the table about a horizontal E-W axis. Restricting the E-W tracking
range of the mirror to $\pm 35^{\circ}$ allows E-W packing of the antennas as
close as 1.3 aperture diameters. The antennas are mounted on the table with
the horn axis at $35^{\circ}$ to the table top; this allows arbitrarily close
packing of the antennas in the N-S direction. Tipping the table by up to
$70^{\circ}$ achieves a zenith angle coverage of $\pm 35^{\circ}$ in the N-S
direction. Most of the tracking motion is individual to the antennas, with
only a small motion of the table (none when observing at the celestial
equator). This results in a fringe rate almost as high as for a fully
individually-mounted array.

The antennas themselves have sidelobe levels of better than $-50$~dB at angles
$> 30^{\circ}$ from the main beam, thanks to careful control
of the boundary conditions (using corrugations) at the edges of the
diffracting aperture. When combined with the shadowing effect of the table and
interferometer effects (fringe rotation and bandwidth smearing) this should
allow daytime operation for all but very close directions to the Sun.

\subsection{Correlator design}

The correlator has to be free from systematics, and insensitive to influences
such as receiver gain instability. A digital corelator is out of the question,
given the bandwidth required ($> 1$~GHz). A simple analogue multiplier, such
as a double-balanced mixer, is sensitive to gain fluctuations in either
input. The design we are using, which was also used in the CAT, is based on
Ryle's phase-switching correlator\cite{ryle}. In Ryle's original design, the
signal from one antenna has a $180^{\circ}$-phase switch inserted, and the
signals from the two antennas are then added. The power of the sum signal is
then detected, and the component at the phase-switch freqency amplified. After
integration, only a term proportional to the product of the two signals
remains, and it is relatively insensitive to gain variations in either side.

\begin{figure}
\psfig{figure=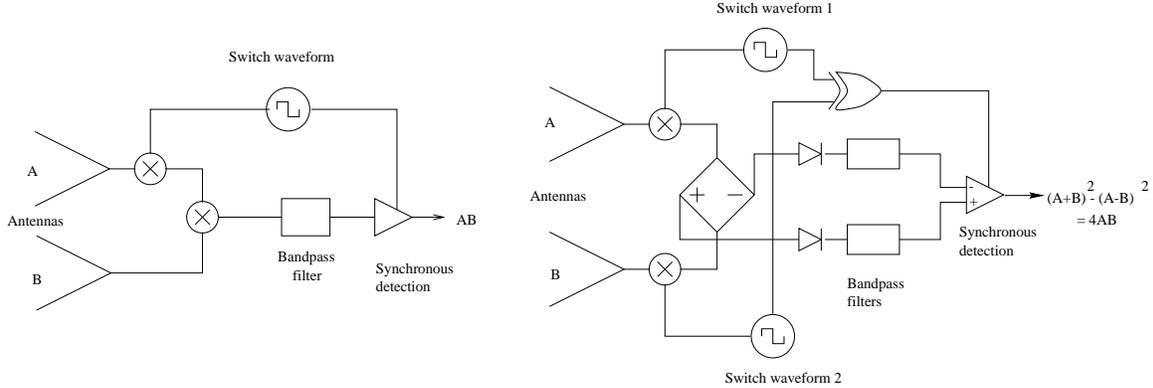,width=6in}
\caption{(Left) Simple multiplying interferometer.  (Right) Plus-minus
correlator as used in the VSA. This is insensitive gain fluctuations in either
input.\label{corr}}

\end{figure}

In the VSA correlator (Fig. \ref{corr}), both the sum and difference are
squared, and their difference taken, before detection at the switch
frequency. This directly removes fluctuations in the total power of each
antenna, providing a further protection against low-frequency noise due to the
receivers or the atmosphere.

This correlator design has excellent performance, but is expensive and bulky,
especially given that we further protect against systematics by housing each
correlator unit in a separate earthed box. This means that we can only afford
to build one frequency channel. However, we need to observe at more than one
frequency in order to separate the CMB signal from the Galactic signal. In
making such spectral-index observations, it is a great advantage to have
well-matched aperture-plane coverage at each frequency. By observing
sequentially at each frequency, we can re-scale the array each time to provide
identical aperture-plane coverage. This would not be possible with
simultaneous observation in widely-spaced observing bands.

\section{Project status}

\begin{figure}
\center{\epsfig{figure=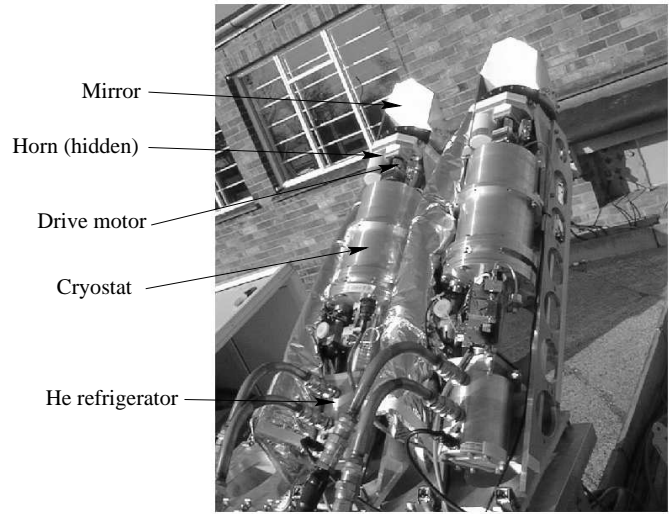,height=3.5in}}
\caption{Two prototype VSA antennas under test.\label{horns}}
\end{figure}

The project was given funding in July 1996, and at the time of writing (March
1998), two complete antennas, with all their associated backend and
correlator, have been built and are being tested (Fig. \ref{horns}). Assembly
of the complete telescope (in Cambridge) will proceed during 1998, and it is
hoped to ship the array to Tenerife during 1999, with science observations
beginning shortly thereafter.

\section*{Acknowledgments}

The VSA is a collaborative project between the Cavendish Laboratory (Cambridge
University), the Nuffield Radio Astronomy Laboratories (Manchester
University), and the Instituto de Astrofisica de Canarias; many people at all
three institutions are responsible for the work described here. The VSA is
funded by the Particle Physics and Astronomy Research Council.

\end{document}